\documentclass[iop,twocolumn,apj]{emulateapj}
\usepackage{natbib,amssymb,amsmath,graphicx,CJKutf8}
\usepackage{hyperref}
\hypersetup{colorlinks=true,linkcolor=magenta,urlcolor=blue,linktoc=page,citecolor=blue}
\shorttitle{}
\shortauthors{Chen, Bastian, \& Gary}
\begin{document}
\title{Direct Evidence of an Eruptive, Filament-Hosting Magnetic Flux Rope Leading to a Fast Solar Coronal Mass Ejection}

\author{Bin Chen \begin{CJK*}{UTF8}{gbsn}(陈彬)\end{CJK*}\altaffilmark{1,3}}

\author{T. S. Bastian\altaffilmark{2}}
\author{D. E. Gary\altaffilmark{1}}
\altaffiltext{1}{Center for Solar-Terrestrial Research, New Jersey Institute of Technology, Newark, NJ 07102, USA}
\altaffiltext{2}{National Radio Astronomy Observatory, Charlottesville, VA 22903, USA}
\altaffiltext{3}{NASA LWS Jack Eddy Fellow; now at: Harvard-Smithsonian Center for Astrophysics, Cambridge, MA 02138, USA; \url{bin.chen@cfa.harvard.edu}}
\begin{abstract}
Magnetic flux ropes (MFRs) are believed to be at the heart of solar coronal mass ejections (CMEs). A well-known example is the prominence cavity in the low corona that sometimes makes up a three-part white-light (WL) CME upon its eruption. Such a system, which is usually observed in quiet-Sun regions, has long been suggested to be the manifestation of an MFR with relatively cool filament material collecting near its bottom. However, observational evidence of eruptive, filament-hosting MFR systems has been elusive for those originating in active regions. By utilizing multi-passband extreme-ultra-violet (EUV) observations from \textit{Solar Dynamics Observatory}/Atmospheric Imaging Assembly, we present direct evidence of an eruptive MFR in the low corona that exhibits a hot envelope and a cooler core; the latter is likely the upper part of a filament that undergoes a partial eruption, which is later observed in the upper corona as the coiled kernel of a fast, WL CME. This MFR-like structure exists more than 1 hr prior to its eruption, and displays successive stages of dynamical evolution, in which both ideal and non-ideal physical processes may be involved. The timing of the MFR kinematics is found to be well correlated with the energy release of the associated long-duration C1.9 flare. We suggest that the long-duration flare is the result of prolonged energy release associated with the vertical current sheet induced by the erupting MFR.

\end{abstract}


\section{Introduction}\label{sect:intro}
Coronal mass ejections (CMEs), typically defined in terms of white-light (WL) solar coronagraphic observations, are the expulsion of coronal plasma and magnetic fields from the lower solar corona into the heliosphere. It has been generally accepted that CMEs are driven by eruptive magnetic flux ropes (MFRs). A typical WL CME often shows a three-part structure, which consists of a leading front, a dark cavity, and a bright core \citep{1985JGR....90..275I}. The leading front is interpreted as compressed coronal plasma ahead of a voluminous MFR, the upper portion of which manifests as the dark cavity \citep[e.g.,][]{2001JGR...10625141L}. The bright core embedded near the bottom of the cavity is often identified as an eruptive prominence \citep{1986JGR....9110951I}.

Similar three-part structures, termed ``prominence cavities'', have been observed to exist in the low quiet-Sun (QS) corona for an extended period of time before they erupt bodily as CMEs \citep{2006ApJ...641..590G}. When viewed at the limb, they appear in WL or EUV images as dark, elliptical structures sometimes hosting a prominence near its bottom. Their disk counterparts are believed to be ``filament channels'' (we will use ``prominence'' and ``filament'' interchangeably in this paper). Both the prominences and the surrounding cavities have been reported to show evidence of helical features, and are believed to share the same magnetic structure as the three-part CMEs: a twisted MFR seen edge on (e.g., \citealt{2006JGRA..11112103G, 2011ApJ...731L...1D}, and references therein). In particular, the concave-upward bottoms of the MFR field lines provide natural support for the dense prominence material, and the upper, density depleted part of the MFR is manifested as the cavity \citep{2000JGR...10518187G,2006JGRA..11112103G,2008ApJ...678..515F, 2009SoPh..256...73V, 2010ApJ...724.1133G}. 

Both the prominences and the surrounding cavities show dynamic behavior, even for non-erupting ones \citep[e.g.,][]{1998Natur.396..440Z, 2003SoPh..212...81K, 2010ApJ...719L.181W, 2013ApJ...770L..28B}. Recently, flows linking prominences and cavities have been reported, which exhibit themselves as horn-like or swirling features \citep[e.g.,][]{2012ApJ...752L..22L, 2013ApJ...770...35S}. They suggest that the prominence and its surrounding cavity is an integrated system in which continuous mass exchange may occur between the two.    

Significant progress in understanding the prominence cavities has been achieved for those in QS regions. Yet observational evidence of such systems remains elusive for those originating in active regions (ARs). This may be attributed to observational difficulties caused by the small spatial and fast dynamical scales of the eruptive MFRs, as well as the complex magnetic field configuration and intense plasma heating in ARs \citep{2013ApJ...764..125P}. Therefore, high-cadence and high-resolution observations sensitive to a wide range of plasma temperatures are necessary to capture the dynamics of fast-evolving MFRs in ARs, which have not been readily available until the launch of the Atmospheric Imaging Assembly (AIA; \citealt{2012SoPh..275...17L}) aboard the \textit{Solar Dynamics Observatory} (\textit{SDO}).  With the \textit{SDO}/AIA observations, a new class of phenomena---hot blobs or ``channels'', depending on the viewing geometry---has been identified as a direct manifestation of MFRs as precursors to fast CMEs originating from ARs \citep{2012NatCo...3E.747Z,2013ApJ...763...43C,2014ApJ...780...28C,2013ApJ...778L..29L,2013ApJ...764..125P}. These eruptive MFR-like hot structures are sometimes enclosed in an expanding dark cavity, or ``bubble'', interpreted to be the main body of the MFR \citep{2011ApJ...732L..25C, 2013ApJ...763...43C, 2012NatCo...3E.747Z}. 

Most of the reported MFR-like hot structures in ARs are only visible in hot AIA channels (i.e., 131 and/or 94 \AA , which are sensitive to $>$6 MK plasma), implying the presence of strong plasma heating. It is suggested that the hot structures are likely heated by magnetic energy release in a current sheet or more generally, a quasi-separatrix layer (QSL) in or around the MFR \citep[e.g.,][]{2013ApJ...778..142T, 2013ApJ...764..125P, 2014ApJ...784...48S}. A similar mechanism is also considered to be the cause of the hot X-ray sheath in QS prominence cavities \citep{2006ApJ...641L.149F, 2012ApJ...758...60F}. However, there has been little direct observational evidence presented of such an MFR-like hot structure that hosts a cooler prominence near its bottom, as an analogue of the QS cavities, in which a prominence is sometimes seen to be located immediately below a hot, X-ray bright sheath inside the cavity \citep{1999ApJ...513L..83H,2012ApJ...746..146R}. We note, however, the recent report by \citet{2014ApJ...789L..35C} that investigates the relationship between a hot channel and the associated prominence material for a ``failed eruption'' event; i.e., no CME was produced. 

Another outstanding question about CMEs is whether they are driven by a pre-existing MFR through \textit{ideal} processes, such as loss of equilibrium or magnetohydrodynamic (MHD) instabilities, or by \textit{non-ideal} (resistive) processes, i.e., magnetic reconnections, which either lead to expulsion of the pre-existing MFR or to the formation of an MFR on-the-fly during the eruption (see, e.g., \citealt{2013AdSpR..51.1967S} and references therein). An alternative that is intermediate to these two scenarios is the ``partial eruption'' model of an MFR, which was first described by \citet{2000ApJ...537..503G} and further developed by \citet{2006ApJ...637L..65G} using MHD simulations. This model invokes both ideal and non-ideal processes in driving the eruption of an MFR: A twisted MFR exists prior to the onset of the CME, which is brought to eruption by an ideal MHD instability. The fast eruption causes the MFR to break into separate parts owing to magnetic reconnection, which may serve as a further driver of the eruption \citep[see the illustration in][]{2001ApJ...549.1221G}. Observational evidence supporting the partial eruption scenario has been reported for several CME events associated with filament eruptions, demonstrating it as a viable means of understanding MFR/CME dynamics for some events \citep[e.g.,][]{2007ApJ...661.1260L,2012ApJ...756...59L,2009A&A...498..295T,2013ApJ...778..142T}. 

In the present paper, using \textit{SDO}/AIA observations, we report direct evidence of a pre-existing, filament-hosting MFR-like structure in the low AR corona which erupts and results in a fast CME and long-duration flare. We adopt a differential emission measure (DEM) analysis method and confirm that the structure consists of a hot envelope and a cooler, filament-like core embedded near its bottom. We then demonstrate that a sudden magnetic reconnection involved in a partial eruption may have facilitated the expulsion of the MFR-like structure into the upper corona. The relevant observations and data analysis are presented in Section \ref{sect:observation}. In Section \ref{sect:discussion}, we briefly summarize the observations and discuss the physical processes involved in the MFR eruption and flare energy release. We then conclude in Section \ref{sect:conclusion}. 

This paper provides the physical framework for follow-up studies of decimetric radio bursts observed in the same event by the Karl G. Jansky Very Large Array (VLA). In particular, this paper sets the stage for detailed analyses of dynamic imaging spectroscopic observations of a variety of decimetric bursts during the energy release phase of this event.

\section{Observations and Analysis}\label{sect:observation}

\subsection{Instrumentation and Data}\label{sect:instrumentation}
The event under study occurred on 2012 March 3 in NOAA AR 11429 on the northeast solar limb. It produced a fast CME and a long-duration soft X-ray (SXR) flare (\textit{GOES} class C1.9) lasting for more than 8 hr. 

Our primary data source is the \textit{SDO}/AIA, which observes the full solar disk up to 1.3 $R_{\odot}$, with a 0$''$.6 pixel size and a 12 s cadence in multiple UV/EUV passbands. In this paper, we use all seven of the EUV passbands, i.e., 304, 171, 193, 211, 335, 94, and 131 \AA , which are sensitive to a wide range of temperatures, from $\sim$0.05 MK to $\sim$11 MK (see \citealt{2010A&A...521A..21O} for details). 

At the time of the present event, one of the two \textit{Solar Terrestrial Relations Observatory} (\textit{STEREO-B}) satellites that trails the Earth's orbit (\textit{STEREO-B}) was separated by 118$^{\circ}$ from the Earth. The location of  allows the same event seen by \textit{SDO} at the east limb to be viewed from another vantage point, from which the AR is located at $\sim$45$^{\circ}$ west of the central meridian, providing a view of the AR against the disk, and a view of the eruption near the limb. Together with the AIA observations, the EUV observations by the Extreme Ultraviolet Imager (EUVI; \citealt{2004SPIE.5171..111W}) aboard \textit{STEREO} (covering 0--1.7 $R_{\odot}$) are used to infer the three-dimensional (3D) context of the AR and the eruption. WL observations from the inner coronagraph COR1 \citep{2003SPIE.4853....1T} aboard \textit{STEREO} (covering 1.5--4 $R_{\odot}$) are used to track the CME in the high corona. 

\begin{figure*}[ht]
\begin{center}
  \includegraphics[width=0.6\textwidth]{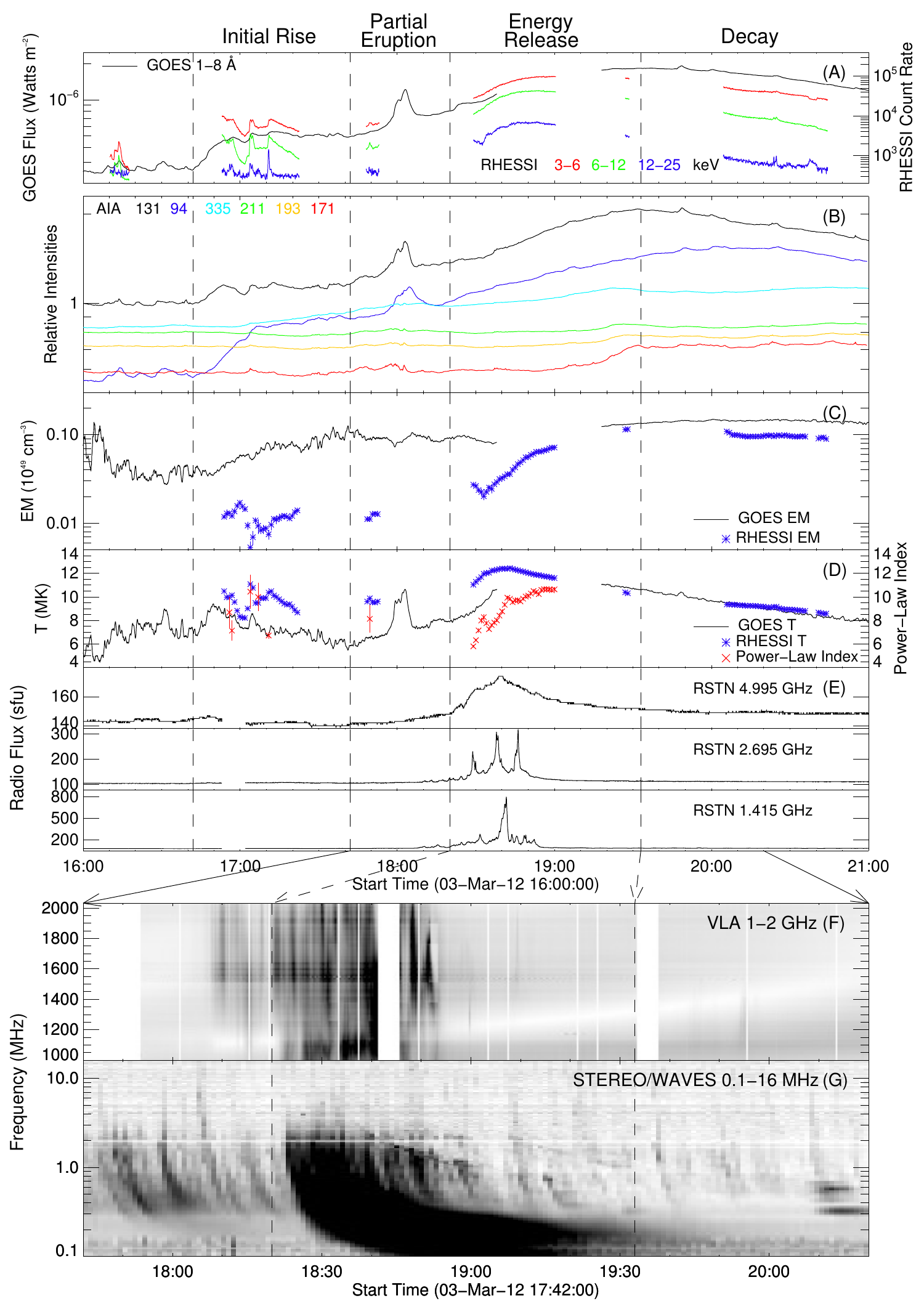}
  \caption{(A) \textit{GOES} (black) and \textit{RHESSI} (color) X-ray light curves. Note they are on different intensity scales. (B) \textit{SDO}/AIA EUV light curves integrated over the entire AR. Each is normalized by the intensity before the event, and then vertically shifted to avoid overlap. The curves are organized from the top to bottom in progressively cooler passbands. (C) Emission measure and (D) temperature of the flare plasma derived from \textit{RHESSI} X-ray spectra (blue) and \textit{GOES} filter ratios (black). Power-law index of the non-thermal X-ray component (red) is also shown in (D). (E) \textit{RSTN} radio flux at 1.415, 2.695 and 4.995 GHz. Note the different $Y$-axis scales. (F) VLA cross-power dynamic spectrum at 1--2 GHz. The data gaps at around 18:43 and 19:35 are due to calibrator scans. (G) \textit{STEREO-B}/WAVES dynamic spectrum at 0.1--16 MHz.}\label{fig:obs_summary}
\end{center}
\end{figure*}

The event was also well-observed by a suite of instruments operating at different wavelengths, including the \textit{Reuven Ramaty High Energy Solar Spectroscopic Imager} (\textit{RHESSI}; \citealt{2002SoPh..210....3L}), \textit{GOES}, the \textit{Solar and Heliospheric Observatory} (\textit{SOHO}), \textit{Hinode}, \textit{WIND}/WAVES, the \textit{Radio Solar Telescope Network} (\textit{RSTN}), and the VLA \citep{2011ApJ...739L...1P}. \textit{RHESSI} observed several time intervals of this event (Figure \ref{fig:obs_summary}(A)), which offer X-ray imaging and spectroscopy above 3 keV, providing information on thermal plasma heated to more than $\sim$8 MK and non-thermal electrons accelerated to energies above $\sim$10 keV. The VLA observed the event from 17:53:23 to 21:44:54 UT with high temporal and spectral resolution (50 ms and 1 MHz) from 1 to 2 GHz. \textit{RSTN} obtained flux density measurements at eight discrete radio frequencies from 245 MHz to 15.4 GHz. \textit{WIND}/WAVES and \textit{STEREO}/WAVES provided dynamic spectral observation of low frequency radio emission from 0.02 to 13.825 MHz and 0.01 to 16.075 MHz. As noted in the introduction, detailed studies of the radio observations and their analysis will be presented in separate publications.  

\subsection{Overview of the Event}\label{sect:overview}

Figure \ref{fig:obs_summary} shows the time history of this event. The \textit{GOES} 1--8 \AA\ SXR flux has an initial rise from the B2.6 level at 16:42 UT to the B5.3 level at 17:05 UT, and then shows a gradual rise until $\sim$17:42 UT, when it increases more rapidly to a short-duration C1.2 peak at 18:03 UT. The SXR flux then declines and resumes its gradual rise until more than 1 hr later at $\sim$19:33 UT when it reaches the maximum SXR level of C1.9. The SXR light curve then shows a slow, several-hour-long decay. Note there is a gap in the \textit{GOES} data between 18:40 and 19:17 UT. 

\textit{RHESSI} only detects X-ray emission at energies below $\sim$25 keV, indicating that the production of high-energy particles is relatively insignificant. \textit{RHESSI} did not observe the C1.2 SXR peak but it captures the main rise phase of the gradual C1.9 flare. The \textit{RHESSI} light curves of the C1.9 flare are relatively smooth with no sign of an impulsive hard X-ray (HXR) peak, although the \textit{RSTN} microwave light curve at 4.995 GHz ($\lambda\approx 6$ cm) shows a peak during the gradual SXR rise, with a maximum of 35 solar flux units (sfu) at 18:39 UT. Consistent with the absence of HXR emission $>$25 keV, no microwave emission is detected in the \textit{RSTN} data at frequencies of 8.8 GHz and higher. In comparison, the \textit{RSTN} light curves in the decimeter wavelength range (2.695 GHz and below) display multiple impulsive peaks. In the VLA dynamic spectrum at 1--2 GHz, these impulsive peaks appear as multitudes of quasi-periodic, short-duration bursts with complex sub-structures. The radio flux density reaches 800 sfu at 1.415 GHz at $\sim$18:41 UT. This event is also associated with a group of interplanetary (IP) type III radio bursts that began at 18:22 UT and coincided with the microwave and decimetric emission. Both the decimetric bursts and the IP type III radio bursts imply the production of non-thermal electrons during the C1.9 flare rise phase. 
   
The associated fast CME ($>$1000 km s$^{-1}$) first appears in the \textit{STEREO}/COR1 field of view (FOV) at 18:31 UT and in the Large Angle Spectrometric Coronagraph (LASCO; aboard the \textit{SOHO} satellite) FOV at 18:48 UT. The inferred CME onset time obtained by extrapolating the quadratic fit of the CME height as a function of time back to the solar surface is $\sim$18:20 UT (see the LASCO CME catalog at \url{http://cdaw.gsfc.nasa.gov/CME_list/}), which coincides with the IP type III burst onset time. 

To guide the following discussion we define four chronologically ordered phases according to the flare/CME evolution (demarcated by the vertical dashed lines in Figure \ref{fig:obs_summary}). They are: (1) the \textit{initial rise} phase (16:42--17:42 UT), for the interval from the event onset to the beginning of the C1.9 gradual flare, (2) the \textit{partial eruption} phase (17:42--18:20 UT), during which the impulsive C1.2 SXR peak occurs, (3) the \textit{energy release} phase (18:20--19:33 UT), covering the main rise of the C1.9 gradual flare until the SXR maximum, and (4) the \textit{decay phase} (19:33 UT and after). We will focus primarily on the first three phases because they involve the most dynamical processes of the eruptive MFR and the flare energy release.

\begin{figure*}
\begin{center}
  \includegraphics[width=1.\textwidth]{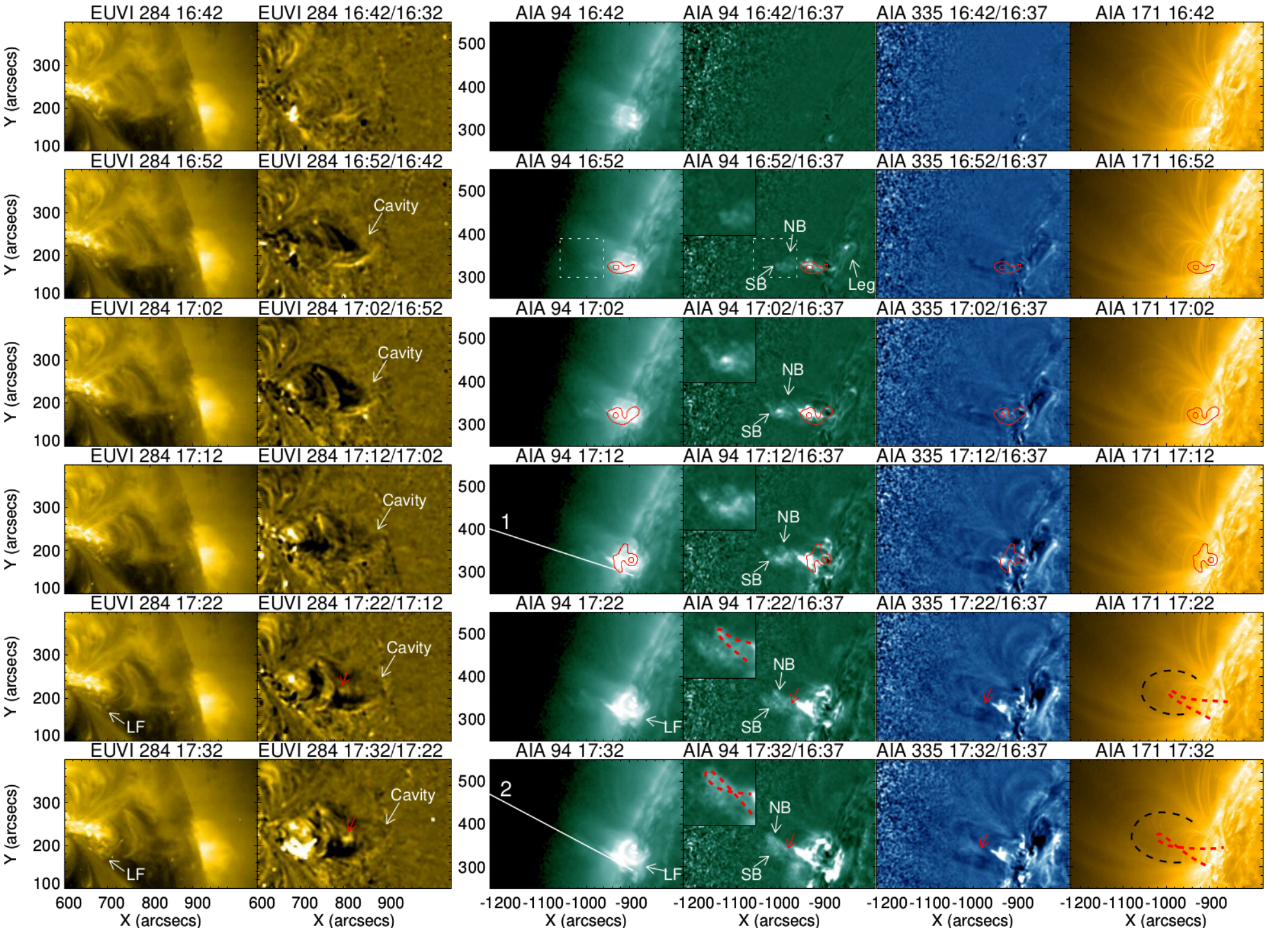}
  \caption{Initial rise phase of the event as seen by \textit{SDO}/AIA and \textit{STEREO}/EUVI-B. Columns 1 and 2 are EUVI-B 284 \AA\ total intensity and running-ratio image sequences (obtained by dividing an image frame by its previous neighbor). Columns 3 and 6 are AIA 94 and 171 \AA\ total intensity image sequences. Columns 4 and 5 are AIA 335 and 211 \AA\ base-ratio image sequences (obtained by dividing an image frame by a pre-event frame at 16:37 UT). Each row shows snapshots at about the same time. The quasi-stationary southern blob-like structure and the eruptive northern helical structure are marked as ``SB'' and ``NB'', respectively. Insets in Column 4 are enlargement of the area indicated by the dashed box in the second row, showing in detail the $\Lambda$-shape to inverted-$\gamma$-shape transition of the eruptive helical structure ``NB''. Red arrows mark the position of the cool core of the upper helical structure, and white arrows mark the low-lying filament (``LF''). Red contours in Columns 3--6 are the \textit{RHESSI} 3--10 keV X-ray sources (contour levels: 30\% and 80\%). Red and black curves outline the kinking helical structure and the surrounding cavity, respectively. Animations of the corresponding EUVI-B and AIA observations are available online.}\label{fig:rise}
\end{center}
\end{figure*}  

\begin{figure*}
\begin{center}
  \includegraphics[width=1.\textwidth]{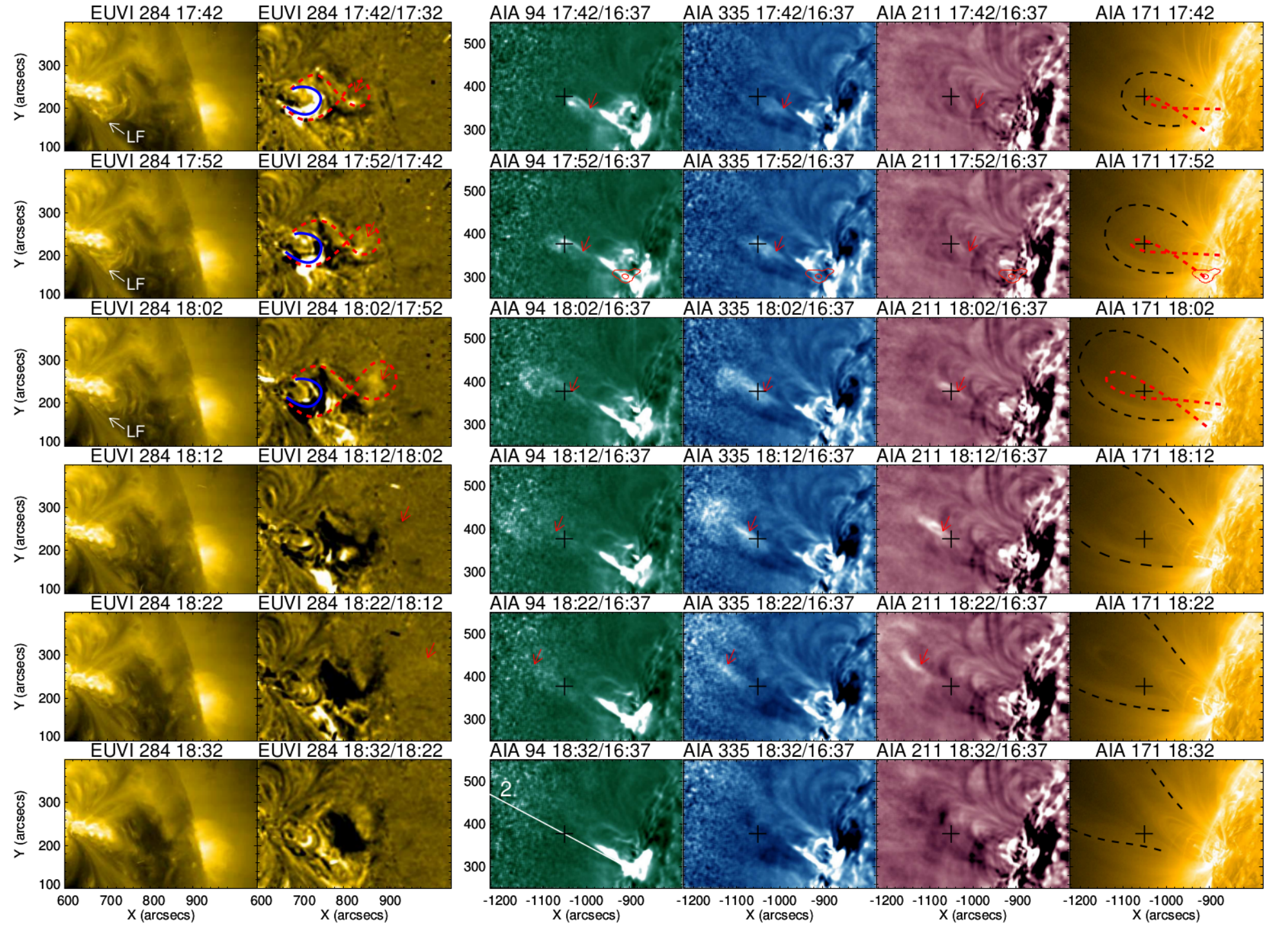}
  \caption{Eruption of the helical structure observed by \textit{SDO}/AIA and \textit{STEREO}/EUVI-B. Columns 1 and 2 are EUVI-B 284 \AA\ total intensity and running-ratio image sequences. Columns 3--5 are AIA 94, 335, and 211 \AA\ base-ratio image sequences (reference frame is at 16:37 UT). Column 6 is AIA 171 \AA\ image sequence. Red contours in Columns 3--6 are the \textit{RHESSI} 3--10 keV X-ray sources (contour levels: 30\% and 80\%). Plus symbol in the AIA images is the region used for the DEM analysis described in Section \ref{sect:eruption}. The low-lying filament is shown by the white arrows. The cool core of the erupting helical structure is indicated by the red arrows in both EUVI-B and AIA images. Red, blue, and black curves outline respectively the erupting helical structure, the low-lying filament, and the surrounding cavity. Animations of the corresponding EUVI-B and AIA observations are available online.}\label{fig:eruption}
\end{center}
\end{figure*}

\subsection{Initial Rise}\label{sect:rise}
The event onset is characterized by the appearance of a hot blob-like structure in AR 11429 above the northeast limb observed by AIA (marked as ``SB'' in Figure \ref{fig:rise}). The blob is seen only in the hot AIA 94 and 131 \AA\ passbands in emission (peak temperature response is 7.1 MK and 11 MK respectively; Figure \ref{fig:rise}, Columns 3--5). At the same time, hot cusp-shaped loops, again only visible in the hot AIA passbands, are formed underneath the blob. The \textit{RHESSI} X-ray light curve shows several impulsive peaks. Imaging and spectroscopy indicate that they are produced by a footpoint source with a rather soft non-thermal component (c.f. Figure \ref{fig:obs_summary}(D)). A more persistent thermal coronal X-ray source is also present, which is co-spatial with the EUV cusp loops with an inferred temperature of $\sim$8--10 MK (rows 2 and 3 of Figure \ref{fig:rise}).  

The blob-like structure remains quasi-stationary in height and disappears in the AIA 94 \AA\ image after $\sim$17:50 UT. No WL CME occurs during this phase, resembling features of a ``failed eruption'' \citep[e.g.,][]{2013ApJ...764..125P}. Another hot structure becomes visible in the AIA 94 \AA\ image beginning at $\sim$17:00 UT to the north of the previous blob (denoted as ``NB'' in Figure \ref{fig:rise}, Column 4). This hot structure has a loop-like shape, which is referred to in the literature as a ``hot channel''  \citep[e.g.,][]{2012NatCo...3E.747Z}. Unlike the quasi-stationary southern blob, this structure maintains a slow, rising motion and at the same time displays a transition from a ``$\Lambda$''-shaped pattern, to a structure showing a closed loop above crossed legs, or an ``inverted-$\gamma$'' pattern (Figure \ref{fig:rise}, Column 4; see also the online animation). Such a morphological evolution is characteristic of kinking of a twisted MFR---the twist is converted to writhe about its main axis \citep[e.g.,][]{2005ApJ...630L..97T}. As viewed from AIA, the leg connecting to the northern footpoint with positive polarity (referred to as the ``northern leg'' hereafter) is located \textit{behind} the one connecting to the southern footpoint with negative polarity (``southern leg''), indicating a left-handed writhe of the helical structure (best seen in Figure \ref{fig:eruption}, Column 3, rows 1--2). No counterpart of this helical structure can be found in cooler AIA passbands than 94 \AA , until $\sim$17:22 UT when a similar feature becomes visible in the AIA 335 \AA\ image.  

As viewed by \textit{STEREO-B}/EUVI at 284 \AA\ (the peak temperature response is $\sim$2 MK, similar to AIA 335 \AA ), the event onset appears as a rising and expanding cavity (Figure \ref{fig:rise}, Column 2). A helical structure does not appear inside the cavity until $\sim$17:22 UT, coinciding in time with the AIA 335 \AA\ observation (Figure \ref{fig:rise}, Columns 1 and 2, rows 5 and 6, indicated by red arrows). From EUVI-B's perspective, this structure has its northern leg \textit{in front of} the southern leg, opposite to the AIA observation (Figure \ref{fig:rise}, Column 1). This is consistent with the \textit{same} structure with left-handed writhe viewing from opposite sides by AIA and EUVI. We note that the early absence of the hot structure in EUVI image is likely due to the lack of a passband that is sensitive to high temperatures (as 131 and 94 \AA\ on AIA).

\begin{figure*}
\begin{center}
  \includegraphics[width=0.7\textwidth]{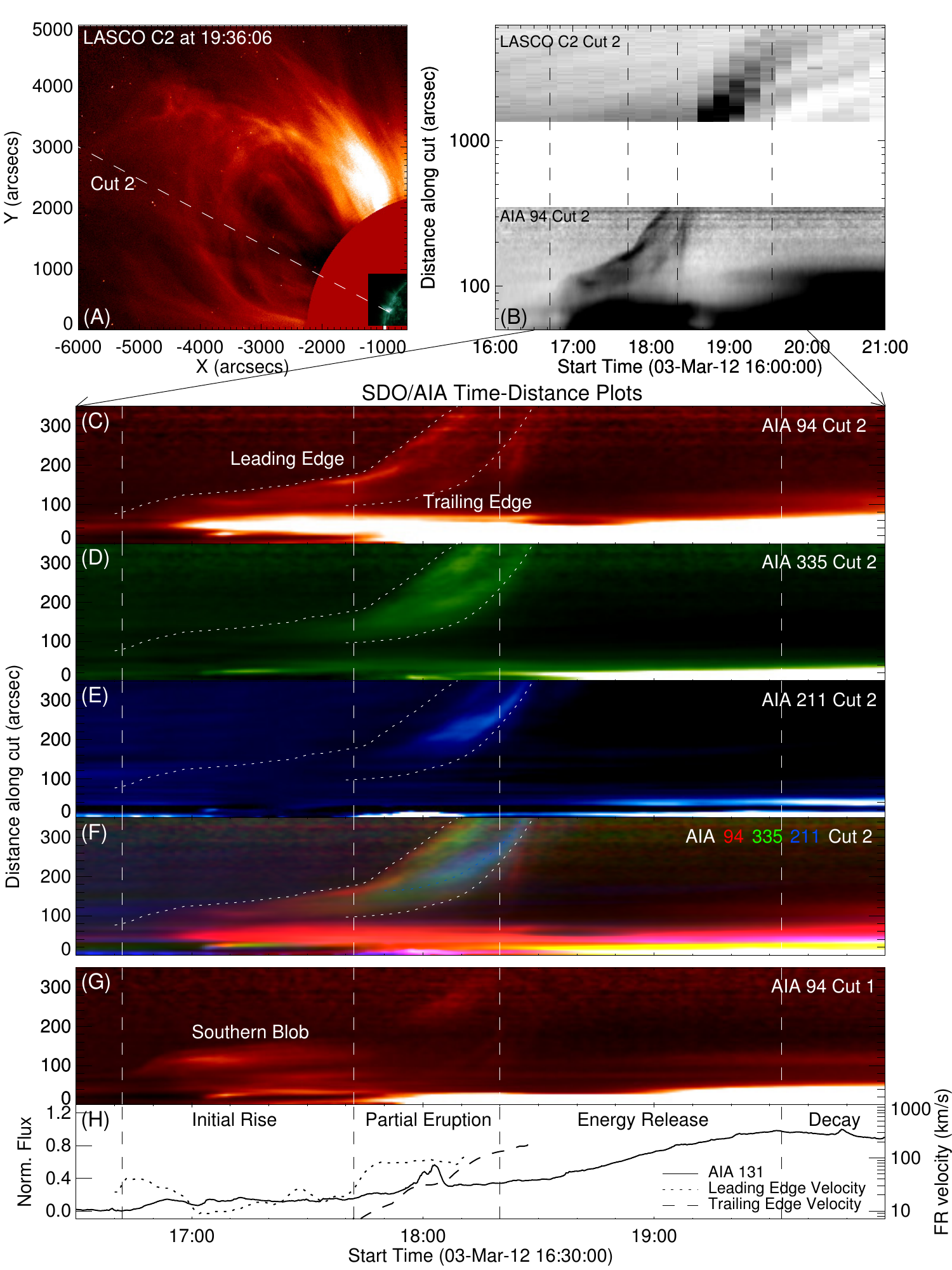}
  \caption{Space--time plots showing the evolution of the MFR-like structures. (A) Composite of AIA 94 \AA\ and LASCO C2 images. Dashed line is along the same direction as Cut 2 as in Figures \ref{fig:rise} and \ref{fig:eruption}. (B) Space--time plot of AIA 94 \AA\ (lower portion) and LASCO C2 (upper-portion). Note the height axis is in logarithm scale. ((C)--(E)) Space--time plots of AIA 94, 335, 211 \AA\ passbands along Cut 2 for the erupting helical structure ``NB''. (F) composite-color plot of all the above three passbands, in which intensities are colored from red to blue for passbands sensitive to decreasing temperatures, showing the spatially distinct multi-thermal structure. (G) AIA 94 \AA\ space--time plot for Cut 1 for the quasi-stationary blob ``SB''.  (H) Leading (dot line) and trailing edge (dashed line) velocity of the eruption. Solid line is the AIA 131 \AA\ intensity integrated over the entire AR. Vertical dashed lines demarcate the four major phases of the event (see Section \ref{sect:overview}).}\label{fig:aia_lasco_t_dist}
\end{center}
\end{figure*}

To better characterize the kinematics of the MFR-like structures, we construct ``space--time'' plots along two cuts in AIA image series (Figure \ref{fig:aia_lasco_t_dist}). Cuts 1 and 2 are positioned along the direction of the southern blob and northern helical structure, respectively (straight white lines in Figures \ref{fig:rise} and \ref{fig:eruption}). Each cut has a width of 5$''$ at the base, which increases linearly with heights corresponding to an angular expansion of 3$^{\circ}$. At any time and give position on the cuts, we average the pixels perpendicular to the cuts within the corresponding width and obtain an intensity value, which makes up a pixel on the space--time plots.

In Figure \ref{fig:aia_lasco_t_dist}(G), the southern blob is clearly seen as a nearly horizontal (stationary) feature with a gradual intensity evolution. The northern helical structure, in contrast, continuously moves to the upper-right direction, indicating a slow, rising motion from $\sim$17:00 UT, followed by a sudden eruption after $\sim$17:42 UT (Figure \ref{fig:aia_lasco_t_dist}(C)). The eruption exits the AIA FOV at $\sim$18:30 UT and evolves coherently as a WL CME in the LASCO/C2 FOV (Figure \ref{fig:aia_lasco_t_dist}(B)). For the remainder of this section we will focus on the northern helical structure because of its major role in the present event. The height--time profile of the erupting helical structure can be conveniently obtained using the space-time plot, shown as the dotted lines in Figures \ref{fig:aia_lasco_t_dist}(C)--(F) for its leading (top) and trailing (bottom) edges respectively. The measurements are made at 1-min intervals and are cross-checked against the intensity versus height profile at any instant. We assign a conservative error of 2$''$ (three AIA pixels) to each measurement. The height--time measurements are smoothed in time before calculating their time derivative to obtain the velocity--time profiles (Figure \ref{fig:aia_lasco_t_dist}(H)). These profiles show that the leading edge of the helical structure first rises slowly at $\sim$20 km s$^{-1}$ until 17:42 UT, when the speed quickly increases to $\sim$100 km s$^{-1}$, whereas its trailing edge stays quasi-stationary and remains in close contact with the underlying cusp loops. 

\begin{figure*}
\begin{center}
  \includegraphics[width=0.8\textwidth]{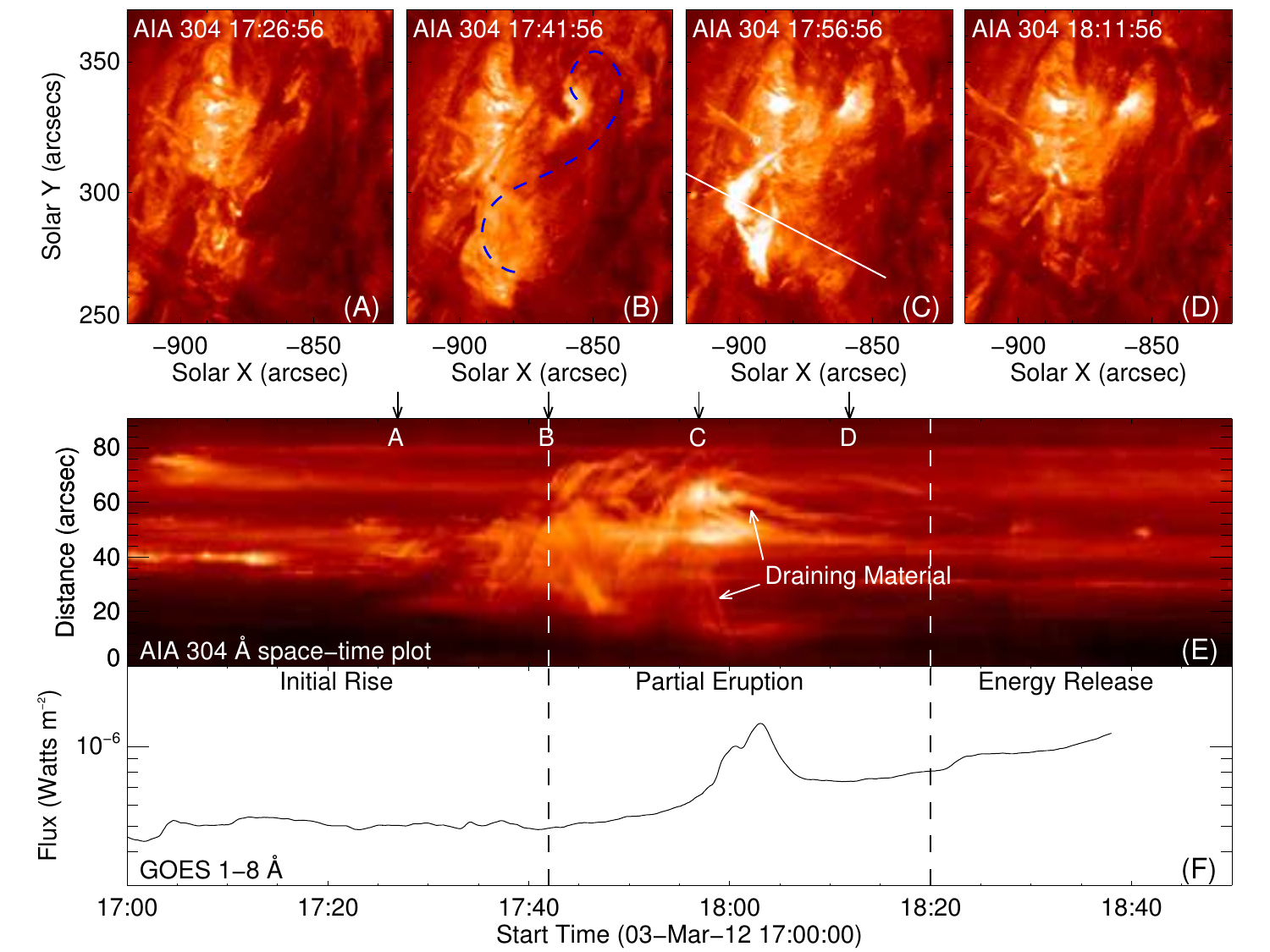}
  \caption{AIA 304 \AA\ observation of the low-lying filament. ((A)--(D)) Snapshots at selected times marked sequentially by the black vertical arrows in (E), the space--time plot obtained along a cut across the filament from 17:00 to 18:40 UT. The cut (marked by the inclined white line) has the same orientation as Cut 2 in Figures \ref{fig:rise}--\ref{fig:aia_lasco_t_dist}. Draining of the filament material begins at $\sim$17:57 UT (white arrows), coinciding in time with the fast rise of the impulsive C1.2 SXR peak, as shown by the \textit{GOES} 1--8 \AA\ light curve (F). An animation of this process is available online.}\label{fig:lf}
\end{center}
\end{figure*}

Another key feature is a low-lying filament that first appears at $\sim$17:22 UT, coinciding with the moment when the eruptive helical structure is first seen in AIA 335 \AA\ and EUVI-B 284 \AA\ images (labeled ``LF'' in Figure \ref{fig:rise}). It is visible across almost all EUVI-B and AIA passbands, illustrating its multi-thermal nature. Viewed from EUVI-B, this filament is nearly face-on with a semi-round shape inclining towards the east. From AIA's perspective, the filament is viewed nearly end-on. The low-lying filament exhibits a reverse $S$-shape in AIA (best seen in Figure \ref{fig:lf}(B)), suggestive of a left-handed writhe \citep{2010A&A...516A..49T}, the same as the eruptive helical structure. The nature of the filament as a twisted MFR is further supported by the continuous helical motion of its many thin threads (see the online animation accompanying Figure \ref{fig:lf}). 

\subsection{Partial Eruption}\label{sect:eruption}
The system enters a new phase at $\sim$17:42 UT, when the previously slowly-rising helical structure starts to ascend quickly. Starting from $\sim$17:55 UT the structure also becomes visible in cooler AIA bands (211 and 193 \AA ). Although the structure shares the similar inverted-$\gamma$ pattern across different passbands, at any given instant, the apex of the helical structure \textit{decreases} in altitude in AIA passbands with descending temperature response (Figure \ref{fig:eruption}, which shows co-temporal snapshot images of AIA 94, 335, 211, and 171 \AA\ from left to right). The hot AIA 94 \AA\ band (7.1 MK) displays a rather sharp edge of the outer ``envelope'', while the warm AIA 335 \AA\ band (2.8 MK) shows diffuse emission filling the 94 \AA\ envelope and the cooler 211 \AA\ (2 MK) features a bright core located near the bottom of the helical structure. The 171 \AA\ image sequence, in contrast, shows a dark cavity surrounding the helical structure. This spatially distinct multi-thermal appearance of the helical structure is also well demonstrated by the composite-color space--time plot in Figure \ref{fig:aia_lasco_t_dist}(F), in which the helical structure seen at 94, 335, and 211 \AA\ is colored from red to blue in descending temperature response. 

\begin{figure*}
\begin{center}
  \includegraphics[width=0.9\textwidth]{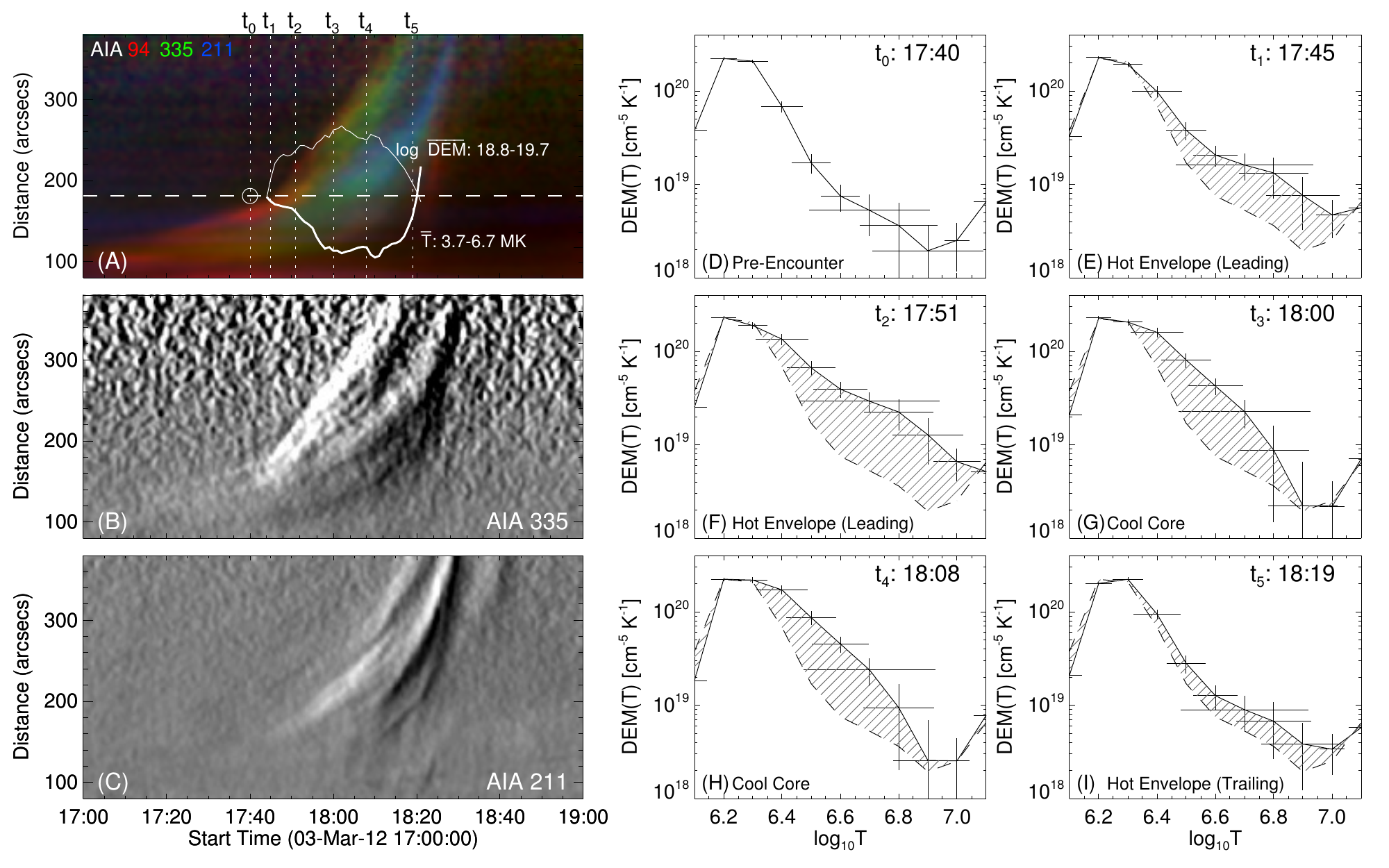}
  \caption{Spatially distinct multi-thermal structures of the eruption. (A) Composite-color space--time plot of the eruption same as Figure \ref{fig:aia_lasco_t_dist}(F). Horizontal line marks the height of the region used for the DEM analysis (the plus symbols in Figure \ref{fig:eruption}). Vertical lines indicate the times ($t_0$--$t_5$) of the DEM curves in (D)--(I). ((B) and (C)) Running-ratio space--time plots for AIA 335 and 211 \AA . (D) Pre-encounter DEM ($t_0$, 17:40 UT) adopted as the background. ((E)--(I)) Solid lines are DEMs for the sampled instants ($t_1$--$t_5$) during the passage of the eruption. Dashed line is the DEM at $t_0$. Their difference, i.e., the time-varying DEM excess (shaded area), is contributed by different spatial parts of the eruption when traversing the selected region, which is then used to obtain the DEM-weighted temperature ($\overline{T}(t)$) and DEM ($\overline{d\xi/dT}(t)$), shown as the thick and thin solid curves in (A). They demonstrate that the eruption consists of a hot envelope with a relatively cooler and denser core embedded near its bottom.}\label{fig:aia_erupt}
\end{center}
\end{figure*}

In order to gain a more quantitative understanding of the multi-thermal structure, we employ the regularized inversion method developed by \citet{2012A&A...539A.146H} to derive the time-dependent DEM distribution of a fixed region in the corona (5$''\times 5''$ box, marked in Figure \ref{fig:eruption} as a plus symbol) using the multi-passband AIA image sequences. The region is reached by the erupting helical structure at $\sim$17:44 UT and passed by after $\sim$18:23 UT, which corresponds to a horizontal line in the space--time plot of Figure \ref{fig:aia_erupt}(A) at a fixed height of $\sim$180 Mm. As different parts of the helical structure reach the region, a time-series of DEM distribution $d\xi/dT(t)$ is obtained, where $\xi$ is the emission measure defined as $\xi=n_e^2h$, $n_e$ is the plasma density, and $h$ is the column depth. Note that these derived DEMs have contributions from all plasma along the line of sight (LOS). 

Figure \ref{fig:aia_erupt}(D) shows an example of the derived DEMs for a time just prior to the encounter (``$t_0$'' in Figure \ref{fig:aia_erupt}(A)). This DEM distribution, adopted as the background, has a major peak at $\sim$1.6 MK ($\log_{10}T\approx 6.2$), which is likely contributed by the background AR coronal plasma along the LOS. Figures \ref{fig:aia_erupt}(E)--(I) shows examples of the derived DEMs (solid lines) for five other selected times during the encounter. All of them display an excess over the background DEM (shaded areas). We attribute the time-varying DEM excess to the net contribution from different spatial parts of the erupting helical structure. 

The time-varying DEM excess shows a broad temperature distribution from 1.6 to 10 MK, implying a multi-thermal nature of the erupting helical structure. Yet it is evident that the DEM excess is weighted toward different temperatures at different times. To demonstrate this quantitatively, we obtain DEM-weighted values of temperature $\overline{T}(t)$ and DEM $\overline{d\xi/dT}(t)$ for the time-varying DEM excess during the passage via
\begin{equation}
 \overline{T}(t)=\dfrac{\int T[\log(d\xi/dT_t)-\log(d\xi/dT_{t_0})]d\log(T)}{\int[\log(d\xi/dT_t)-\log(d\xi/dT_{t_0})]d\log(T)},
\end{equation}
\begin{equation}
 \overline{d\xi/dT}(t)=\dfrac{\int d\xi/dT_t[\log(d\xi/dT_t)-\log(d\xi/dT_{t_0})]d\log(T)}{\int[\log(d\xi/dT_t)-\log(d\xi/dT_{t_0})]d\log(T)},
\end{equation}
shown as the solid curves in Figure \ref{fig:aia_erupt}(A). $\overline{T}$ decreases from $\sim$6 MK for the leading edge to $\sim$4 MK for the core, and increases back to $\sim$7 MK for the trailing edge, consistent with the red-blue-red color transition along the horizontal line in Figure \ref{fig:aia_erupt}(A). The weighted DEM profile $\overline{d\xi/dT}$ shows the opposite: it increases from $6\times 10^{18}$ cm$^{-5}$ K$^{-1}$ for the leading edge to $5\times 10^{19}$ cm$^{-5}$ K$^{-1}$ for the core by about an order of magnitude, and decreases again when the trailing edge meets the sampled region. Assuming a single column depth $h$, the core is about a factor of 3 denser than the outer envelope. We emphasize that the values above are based on DEM-weighted and LOS-integrated quantities, and moreover, our DEM analysis does not extend to chromospheric temperatures. In fact, we do see a counterpart in the AIA 304 \AA\ image (peak temperature response is 50,000 K), which aligns well in position with the cool core, implying that the DEM of the core extends to chromospheric temperatures. Therefore, the above temperature and density contrasts derived above between the envelope and the core ($\sim$1.5 and 3, respectively) should only be considered as lower-limits.  

The AIA 335 and 211 \AA\ running-ratio space--time plots (Figures \ref{fig:aia_erupt}(B) and (C)) show that the eruption consists of many spatially distinct, upward-propagating tracks, implying that it is made up of many nested ``shells'', each has a thickness of $\lesssim$5$''$ (3.6 Mm). Evidence of nested, shell-like sub-structures have been observed in quiescent prominence cavities \citep{2010ApJ...719.1362H, 2011A&A...533L...1R, 2013ApJ...770L..28B}, and are also distinguishable in the previously reported CME-associated eruptive ``hot channels''  \citep[e.g.,][]{2014ApJ...780...28C}.  

Starting from $\sim$17:50 UT, the EUVI 284 \AA\ image, in which both the low-temperature portion of the eruptive helical structure and the low-lying filament are readily visible, shows that the two quickly detach from each other (Figure \ref{fig:eruption}, Column 2, rows 1--3; see also the accompanying online animation for a better representation). The core of the helical structure is then expelled toward higher altitudes and becomes the coiled kernel of the WL CME observed in COR1 (see next subsection). At about the same time, the low-lying filament shows a sudden draining of its material towards its two conjugate footpoints and then disappears (also evidenced by the downward- and upward-moving tracks in Figure \ref{fig:lf}(E)). The draining motion of the filament material coincides very well in time with the rapid rise of the short-duration C1.2 SXR peak (Figures \ref{fig:lf}(E) and (F)), suggesting the occurrence of a sudden magnetic reconnection.

The features presented above are consistent with the ``partial eruption'' scenario, in which an internal reconnection leads to bifurcation of an erupting MFR. Noting that the eruptive helical structure and the low-lying filament appear almost synchronously and share the same sense of writhe, we suggest that they represent different branches of the same MFR that formed earlier in this event.

\subsection{Energy Release Phase}\label{sect:impulsive} 

\begin{figure*}
\begin{center}
  \includegraphics[width=1.0\textwidth]{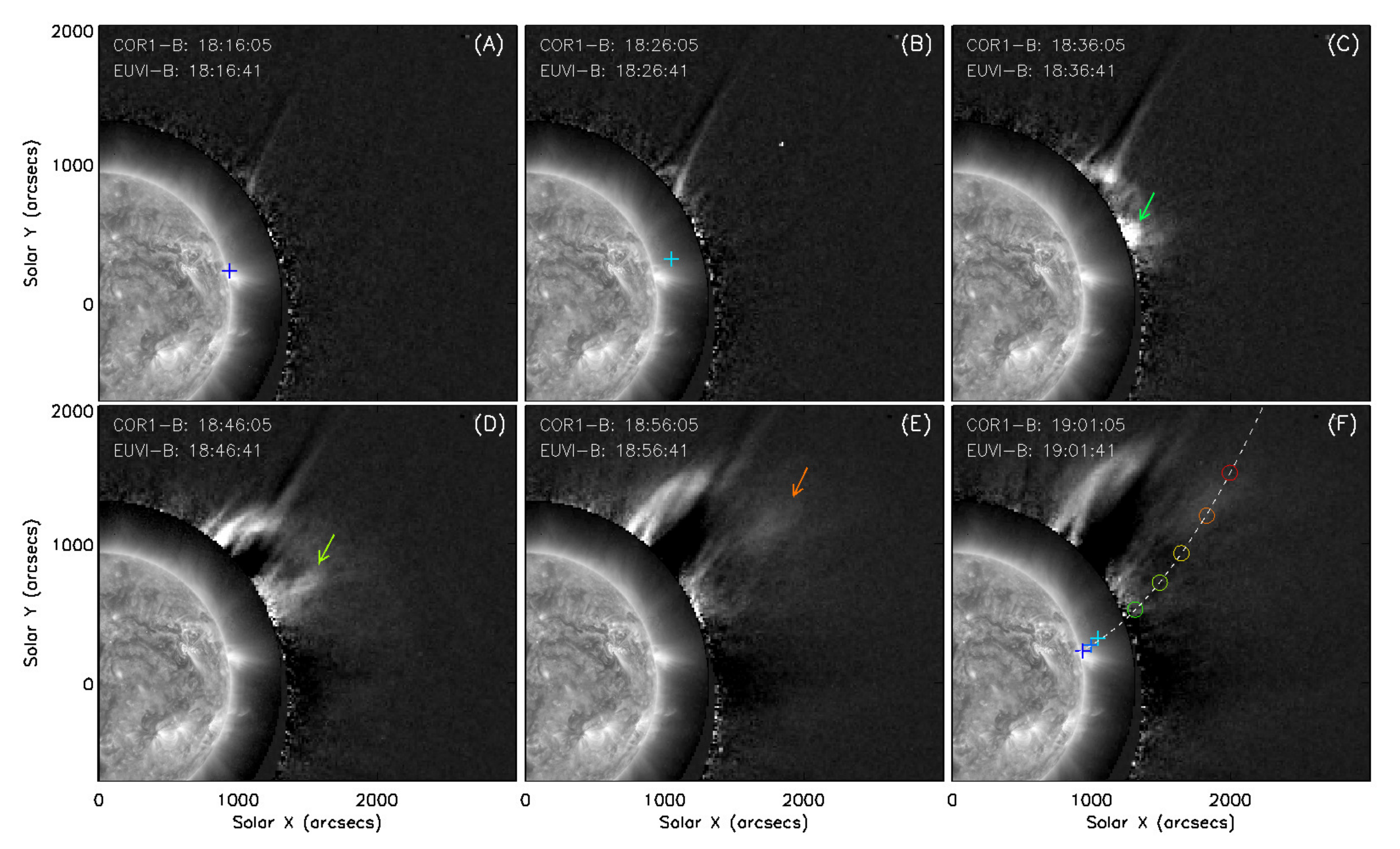}
  \caption{Composite of EUVI-B 171 \AA\ and COR1-B WL images showing the eruption of the filament-embedded-cavity from low to high corona, where it becomes a three-part WL CME with a coiled core, a cavity, and a leading front. Plus symbols in (A) and (B) and arrows in (C)--(E) mark the position of the erupting filament in the low corona, observed by EUVI-B 284 \AA\ as an inverted-$\gamma$-shaped feature, and in the upper corona, where it is manifested as the coiled core of the three-part CME. (F) shows examples of the measured centroid positions of the erupting filament from 18:16--19:01 UT (plus and circular symbols are for EUVI-B 284 \AA\ and COR1-B measurements respectively), colored in progressively later time from blue to red. The filament centroid positions can be well-fitted by a parabolic curve (dashed line), which is used as a guide to obtain the distance--time profile $r(t)$ in Figure \ref{fig:fr_dynm}(A). 
}\label{fig:euvi_cor1_cme}
\end{center}
\end{figure*}

Immediately after the helical structure exits the EUVI-B FOV, from 18:31 UT, a bright front emerges in the COR1 FOV, followed by a coiled core embedded in a dark cavity, exhibiting a three-part WL CME (Figure \ref{fig:euvi_cor1_cme}). The centroid location of the core of the helical structured and the coiled CME kernal can be connected smoothly and fit with a parabolic arc (Figure \ref{fig:euvi_cor1_cme}(F)), suggesting that they are associated with the same eruption. 

\begin{figure*}
\begin{center}
  \includegraphics[width=0.8\textwidth]{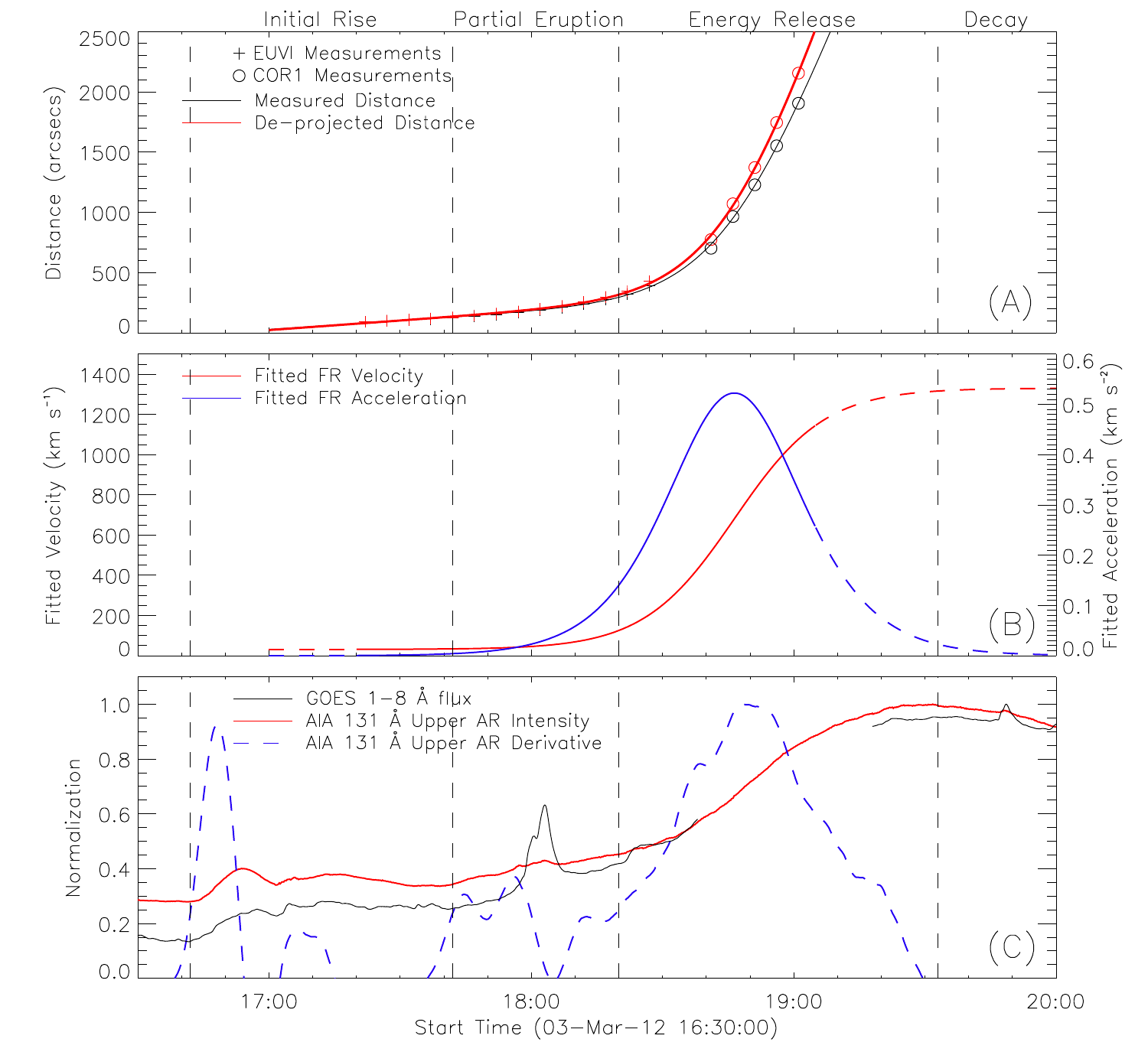}
  \caption{Kinematics of the MFR and its association with the gradual C1.9 flare. (A) Distance--time profile $r(t)$ of the MFR core from the low (plus symbols, from EUVI-B 284 \AA\ measurements) to high corona (circular symbols, from COR1-B measurements). Black and red curves are the fitted results of $r(t)$ based on measured and de-projected distances respectively (see the text for details). (B) Velocity--time ($v(t)$, red) and acceleration-time ($a(t)$, blue) profiles derived from the first- and second-derivatives of the fitted, de-projected, $r(t)$ curve. (C) \textit{GOES} 1--8 \AA\ light curve (black solid line), AIA 131 \AA\ intensity integrated over the upper AR (which effectively removes the contribution of the impulsive C1.2 peak; red solid line), and its time derivative (blue dashed line).}\label{fig:fr_dynm}
\end{center}
\end{figure*}

Using the fitted trajectory as a guide, we obtain the projected distance-time profile $r(t)$ of the MFR from the low to high corona (Figure \ref{fig:fr_dynm}(A), black symbols). We then apply a tie-point triangulation technique based on the simultaneous EUVI-B and AIA image pairs to derive a 3-D trajectory of the eruption, which is used to obtain the de-projected $r(t)$ profile of the erupting MFR in the low corona viewed from \textit{STEREO-B} (Figure \ref{fig:fr_dynm}(A), red symbols). For the COR1 measurements in the high corona where no observation is available from another vantage point, we linearly extrapolate the projection angles outward to correct for the projection. We find that the projection effect is relatively small because the eruption propagates within a small angle away from the plane of sky as viewed from EUVI-B.

We then adopt the convenient functional form given by \citet{2007ApJ...671..926S} \citep[see also][]{2010A&A...522A.100P} 
\begin{equation}
r(t)=r(t_p)+\frac{1}{2}(v_f+v_0)(t-t_p)+\frac{1}{2}(v_f-v_0)\tau\ln\left[\cosh\dfrac{t-t_p}{\tau}\right]
\end{equation}
to fit the de-projected distance--time function $r(t)$ (Figure \ref{fig:fr_dynm}(A), solid curves). Here, $v_0$ and $v_f$ are the initial and terminal velocities, $t_p$ is the time corresponding to the peak acceleration, and $\tau$ is the time scale of the acceleration. This function is selected because it can reproduce the $r(t)$ and velocity--time $v(t)$ profiles ranging from nearly constant acceleration to impulsive acceleration, which have been previously demonstrated to resemble profiles obtained from CME observations and simulations very well \citep{2010A&A...522A.100P}. Then the velocity and acceleration time profiles $v(t)$ and $a(t)$ can be obtained by taking the first and second derivatives of the de-projected $r(t)$ fit (Figure \ref{fig:fr_dynm}(B)). Using this technique, we effectively circumvent the large and sometimes unacceptable uncertainties of the $a(t)$ profile that result from taking the derivatives of $r(t)$ directly. 

The $v(t)$ and $a(t)$ profiles clearly display a slow rise after the cool component of the helical structure appears at 17:22 UT, followed by an acceleration occurring around 18:00 UT, coinciding with the time of the short-duration C1.2 SXR peak and the partial eruption. Then, the helical structure continues to accelerate as it leaves the EUVI FOV and becomes the WL CME core observed in COR1. The maximum acceleration is reached at $\sim$18:43 UT. After that, the acceleration starts to decrease and the CME core maintains a nearly constant velocity of $\sim$1300 km s$^{-1}$ before it becomes too diffuse to be measured.

The \textit{GOES} 1--8 \AA\ SXR flux looks very similar to the velocity profile of the eruption. However, direct comparison between the \textit{GOES} derivative and the acceleration profile is not possible because of the \textit{GOES} data gap in 18:40--19:17 UT. We instead use the AIA 131 \AA\ light curve as a proxy of the \textit{GOES} flux to obtain the temporal derivative. This works reasonably well because both are sensitive to plasma temperatures $\gtrsim$10 MK \citep[e.g.,][]{2011ApJ...739...59W, 2012ApJ...746L...5S, 2013ApJ...764..125P}. Figures \ref{fig:fr_dynm}(B) and (C) show that the resulted AIA 131 \AA\ derivative is closely synchronized with the acceleration profile $a(t)$ of the eruption, which conforms with the well-known synchronization between flare emissions and MFR/CME acceleration \citep[e.g.,][]{2001ApJ...559..452Z}, both interpreted to be driven by magnetic reconnection.

As the temperature sensitivity progresses to cooler plasmas from AIA 131 to 171 \AA , the EUV light curves are sequentially delayed to later times (Figure \ref{fig:obs_summary}(B)). \textit{RHESSI} X-ray flux shows a similar gradual rise but it peaks much earlier than the EUV curves (Figure \ref{fig:obs_summary}(A)). In addition, the \textit{RSTN} microwave emission at 4.995 GHz, presumably produced by gyrosynchrotron radiation from higher-energy non-thermal electrons, reaches the maximum intensity even earlier (at $\sim$18:39 UT). The extended rise and decay of the SXR light curve, together with the trend that their peaks appear sequentially in emissions characteristic of lower temperatures or electron energies, provides clear evidence of an extended heating associated with the rise phase of the gradual C1.9 event (see \citealt{2012ApJ...746L...5S} for another example). The extended heating is also evidenced by the continuously rising plasma temperature derived from the \textit{GOES} filter ratios (Figure \ref{fig:obs_summary}(D)).

The MFR expulsion and the gradual rise of the \textit{GOES}/AIA 131 light curves also coincide in time with the onset of the IP type III burst at $\sim$18:22 UT; the latter is a strong indication of particle acceleration occurring in the ``impulsive phase'' of a flare (see \citealt{2011SSRv..158....5H} for a review), here referred to as the ``energy release'' phase because the standard hallmark of the impulsive phase---the presence of impulsive HXR emission---is not detected. Furthermore, a non-thermal power-law component is found to persist in the X-ray spectrum above 10 keV, imaged to be a loop-top HXR source (not shown here), which serves as another signature of ongoing particle acceleration (see \citealt{2008A&ARv..16..155K} for a review).

\section{Discussion}\label{sect:discussion}

\begin{figure*}
\begin{center}
  \includegraphics[width=0.9\textwidth]{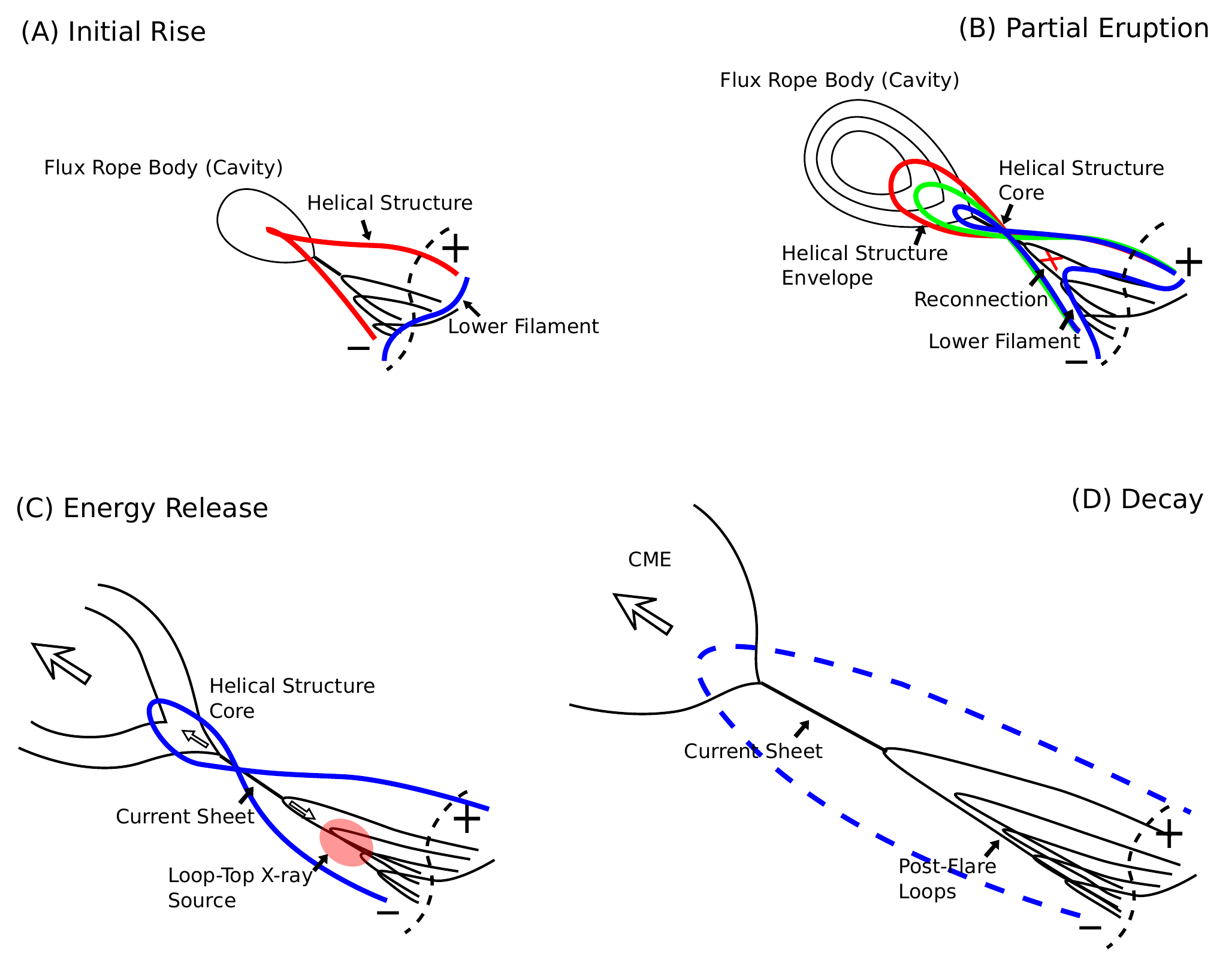}
  \caption{Schematic depiction of the proposed physical scenario in terms of the four phases: (A) \textit{initial rise} phase, when a twisted MFR emerges into the corona. The thick red curve represents the observed helical structure, whereas the thin black oval denotes the MFR cross-section manifested as the surrounding cavity. The blue thick curve denotes the low-lying filament; (B) \textit{partial eruption} phase, in which the helical structure evolves from an initial $\Lambda$-shape to an inverted-$\gamma$ shape and starts its fast rise. It features a hot envelope (red and green) with a cool core (blue) embedded near the bottom. An internal reconnection (denoted as a red $\times$ symbol) occurs and causes the partial eruption; (C) \textit{energy release} phase, in which the expelled helical structure induces a vertical current sheet in which prolonged magnetic reconnection occurs; and (D) \textit{decay} phase, during which the eruption develops into a CME and post-flare loops appear.}\label{fig:cartoon}
\end{center}
\end{figure*}

The timeline of the event is briefly summarized as follows:
\begin{enumerate}
\item{\textit{Initial rise (16:42--17:42 UT).}} A slow-rise, kinking helical structure appears in hot AIA passbands. Later, a low-lying filament appears when the helical structure also becomes visible in cooler AIA passbands, both of which have a left-handed writhe. Episodic X-ray footpoint emission occurs and hot cusp loops develop. 

\item{\textit{Partial eruption (17:42--18:20 UT).}} The helical structure starts to rise quickly. The apex of the same helical structure appears at \textit{decreasing} heights in AIA passbands sensitive to lower temperatures. DEM analysis suggests that the helical structure consists of a hot envelope and a cooler core situated near its bottom. During the short-duration C1.2 peak, the helical structure quickly detaches from the low-lying filament. Meanwhile the low-lying filament material suddenly drains to the solar surface.

\item{\textit{Energy release phase (18:20--19:33 UT).}} The expelled helical structure propagates into the upper corona and appears as the coiled kernel of the three-part WL CME, during which the gradual rise of the C1.9 flare and the main acceleration of the eruption occur. The acceleration profile of the eruption is closely synchronized with the AIA 131 \AA\ derivative. Various phenomena are observed in support of prolonged particle acceleration and heating occurring in this phase.

\item{\textit{Decay phase (19:33 UT and after).}} Myriad cusp-shaped post-flare loops form following the eruption and show shrinking motion from the cusp-tips. They appear sequentially in AIA passbands sensitive to decreasing temperatures. 
\end{enumerate}

\subsection{Proposed Physical Scenario}\label{sect:physics}
Here we incorporate the observed phenomena and propose a physical scenario in terms of the MFR evolution and the associated flare emission. The major elements are depicted in sketches shown in Figure \ref{fig:cartoon} organized by the four major chronological phases.  

\textit{Initial rise}. An MFR arises in the AR corona and appears as a hot helical structure. The exact mechanism of the MFR initialization, however, is beyond the scope of the current study. We refer to the review by \citet{2011LRSP....8....1C} for further discussion of this point. The lower portion of the voluminous MFR, where relatively dense material collects, is manifested as the observed helical structure, while the upper, density-depleted portion of MFR appears as the dark cavity. The ascending MFR stretches the overlying envelope magnetic field lines and leads to the formation of the underlying cusp loops. Bursty magnetic reconnections can occur in a QSL in or around the MFR. The released magnetic energy may be the source of the episodic non-thermal X-ray spikes and heating of the helical structure, cusp loops, and the low-lying filament. The slow rising motion of the helical structure could be the result of ideal processes such as flux emergence and footpoint motions, or non-ideal processes, i.e., reconnections, which add flux and twist to the helical fields of the MFR. 

\textit{Partial eruption}. A helical kink instability may be triggered if the accumulated twist exceeds a critical value \citep{1981GApFD..17..297H}, causing the MFR to rise quickly. The observed morphological evolution and slow-to-fast-rise transition are very similar to the simulation results in \citet{2005ApJ...630..543F} based on a model of a kinked MFR. This points to an ideal process as the initial driver of the MFR fast rise. The erupting, helical branch of the MFR hosts a cooler and denser filament near its bottom; the latter is magnetically connected to another low-lying filament which may belong to the same MFR system. As the erupting filament-MFR system continues its fast rise, internal magnetic reconnection can occur, which causes the division of the helical structure and the low-lying filament. The reconnection results in expulsion of the helical structure, and at the same time, causes the low-lying filament material to lose magnetic support and drop back toward the solar surface. This sudden reconnection is also manifested as the short-duration C1.2 SXR peak occurring at the same moment.

\textit{Energy release phase}. After the helical structure is expelled into the upper corona as a CME, the subsequent evolution is largely consistent with the standard CSHKP flare-CME model \citep{1964NASSP..50..451C, 1966Natur.211..695S, 1974SoPh...34..323H, 1976SoPh...50...85K}: the erupting MFR induces a vertical current sheet behind it in which magnetic reconnection occurs, resulting in the gradual C1.9 flare and the extended MFR acceleration. This is demonstrated by the close synchronization between the MFR acceleration and the AIA 131 \AA\ derivative (as a proxy of the thermal energy release rate). Such a correlation is well-known and has been considered as supporting evidence for the CSHKP model in which reconnection is the common mechanism driving the CMEs and the associated flares \citep[e.g.][]{2004ApJ...604..420Z, 2004SoPh..225..355V, 2004ApJ...604..900Q, 2005ApJ...620.1085J, 2012ApJ...755...44B}. By using numerical modeling, \citet{2006ApJ...644..592R} shows that the Lorentz force from the current sheet and envelope fields can accelerate the MFR, while the Poynting flux in the current sheet gives rise to the flare energy release. Other phenomena observed during this phase, which suggest an extended period of particle acceleration and heating, can all be consequences of the prolonged magnetic reconnection. 

\textit{Decay phase}. As the reconnection proceeds to higher and higher altitudes following the eruption, previously heated loops cool due to conductive and radiative losses, while newly reconnected loops, some of which are observed as rapidly shrinking loops from the cusp tip, are heated by the gradually released magnetic energy. The post-flare loops that appear sequentially in cooler and cooler AIA passbands are also evidence of the extended cooling.

\section{Concluding Remarks}\label{sect:conclusion}
As noted in Section \ref{sect:instrumentation}, the VLA observed this event at decimeter wavelengths from 17:53 UT to well past the SXR maximum. During the rise to maximum---the energy release phase---a rich variety of decimetric radio bursts was recorded with unprecedented spatial, spectral and temporal resolution over a wide frequency band. These observations allow each pixel in the radio dynamic spectrum to be imaged (``dynamic imaging spectroscopy''), enabling new radio diagnostics of magnetic energy release processes and the environment in which they occur. This paper provides the physical context and framework within which detailed analyses of the VLA data will be presented in future publications.

In this paper we report direct observational evidence of a filament-hosting, MFR-like helical structure in the low corona originating from an AR that leads to a fast CME and a long-duration flare. The structure features a hot envelope only visible in the AIA 131, 94 and 335 \AA\ passbands that are sensitive to high temperature plasma, along with a relatively cooler and denser core embedded near its bottom, seen by the cooler AIA 211, 193, 171, and 304 \AA\ passbands. DEM analysis based on the multi-passband AIA EUV data confirms the existence of spatially distinct multi-thermal sub-structures, and reveals that the core is at least 1.5 times cooler and three times denser than the envelope. The MFR-like structure first emerges with a surrounding dark cavity in the low corona and later erupts into the upper corona as a three-part WL CME, with its cool core making up the coiled kernel of the CME. The structure shows a slow- to fast-rise transition prior to its eruption, and in the mean time, evolves from a $\Lambda$-shape to an inverted-$\gamma$ shape, closely resembles the simulation results in \citet{2005ApJ...630..543F} based on a model of an MFR erupting upon helical kink instability. 

We conclude that the eruptive helical structure is the manifestation of the lower portion of a voluminous MFR where relatively dense material collects, while the upper, density-depleted portion of the MFR is seen as the surrounding dark cavity. The helical structure shows a closely synchronized evolution and the same sense of writhe as a lower-lying filament. We attribute the two features to be different branches of the same voluminous MFR, which later undergoes a partial eruption induced by a sudden internal magnetic reconnection. The helical structure is heated by magnetic energy release in a QSL in or around the MFR to high temperatures ($>$6 MK), in which a cooler filament-like core is embedded. This appears very similar to the hot, X-ray bright sheath observed in quiescent cavities which sometimes appears immediately above and around a prominence \citep{1999ApJ...513L..83H,2012ApJ...746..146R}, yet the structure in our case is of course much smaller, hotter and more dynamic because it originates in an AR and leads to a fast CME. The partial eruption process culminates in the expulsion of the helical structure, draining of the low-lying filament material towards the solar surface, and the co-temporal impulsive C1.2 SXR peak. All the major elements of the partial eruption scenario as illustrated by \citet{2000ApJ...537..503G} and predicted by the numerical simulations in \citet{2006ApJ...637L..65G}, including the presence of a filament-hosting MFR, the slow-to-fast-rise transition of the kinking MFR, the sudden reconnection and MFR bifurcation, are observed in this single event thanks to the multi-passband AIA observations.
 
We also demonstrate that the MFR-like structure was present in the low AR corona for at least an hour before the onset of its eruption, suggesting that the eruption is initially driven by an \textit{ideal} MHD process, likely a helical kink instability. However, we argue that the ideal MHD instability is not the sole mechanism driving the eruption, and that a \textit{non-ideal} process, i.e., magnetic reconnection, also plays an important role in facilitating the MFR eruption. This is supported by the onset of the major acceleration phase of the MFR coinciding with the short-duration C1.2 SXR peak, and the extended acceleration period of the MFR/CME later in the energy release phase. The former corresponds to sudden reconnection due to the MFR bifurcation, and the latter is likely associated with the prolonged magnetic reconnection in the vertical current sheet behind the eruption.

Both hot MFR-like structures (blobs/hot channels) and cool filament-like structures originating from ARs have been frequently reported in association with CMEs and flares in the literature. \citet{2014ApJ...789L..35C} recently explored the relationship between the hot channel and the associated filament, finding an evolution similar to that reported here, although the event in question was a ``failed eruption'' that did not produce a CME. The event described here simultaneously shows a hot MFR-like structure, enclosed by a dark cavity, and a relatively cool and dense filament embedded near its bottom. We are able to follow the entire evolution of the filament-MFR ensemble through a partial eruption and energy release associated with the ejection of a fast CME. We note that the spatially-distinct multi-thermal structure of the MFR presented here can only be detected upon careful examination of all the AIA passband images with full resolution. Even so, the fast temporal evolution of the eruption, the small spatial scale of the helical structure, and the low contrast of the hot component against the background make such structures easy to overlook. Therefore, it is likely that similar features may be present in other events but have not been recognized before. It is equally possible that some of the MFR-like structures may not host a filament even if the dipped fields of the MFR are capable of providing the necessary support. The latter is similar to the case of the filament channel, observed against the solar disk as a density-depleted region in which a filament may or may not exist \citep{1998SoPh..182..107M, 2006JGRA..11112103G}. In any case, whether it is merely a special case or a common phenomenon will motivate further investigation of more events like this one.   

\acknowledgements
We thank Chang Liu for help on the data analysis, Iain Hannah for the DEM inversion code, Gregory Fleishman for insightful discussions, and Wei Liu for providing the first AIA movies. The authors are grateful to the teams of \textit{SDO}, \textit{STEREO}, and \textit{RHESSI} for making the data available. This research was supported by the NASA Living With a Star Jack Eddy Postdoctoral Fellowship Program, administrated by the University Corporation for Atmospheric Research. The National Radio Astronomy Observatory is a facility of the National Science Foundation operated under cooperative agreement by Associated Universities, Inc. DG acknowledges support from NSF grant AST-1312802 to NJIT. 


\end{document}